\documentclass[]{aa} 

\usepackage{graphicx}
\usepackage{txfonts}
\usepackage[normalem]{ulem}
\usepackage{color}
\usepackage{esvect}

\definecolor{forestgreen(web)}{rgb}{0.13, 0.55, 0.13}

\usepackage[colorlinks,allcolors=blue]{hyperref}

\newcommand{\galaxy}{\includegraphics[width=0.3cm]{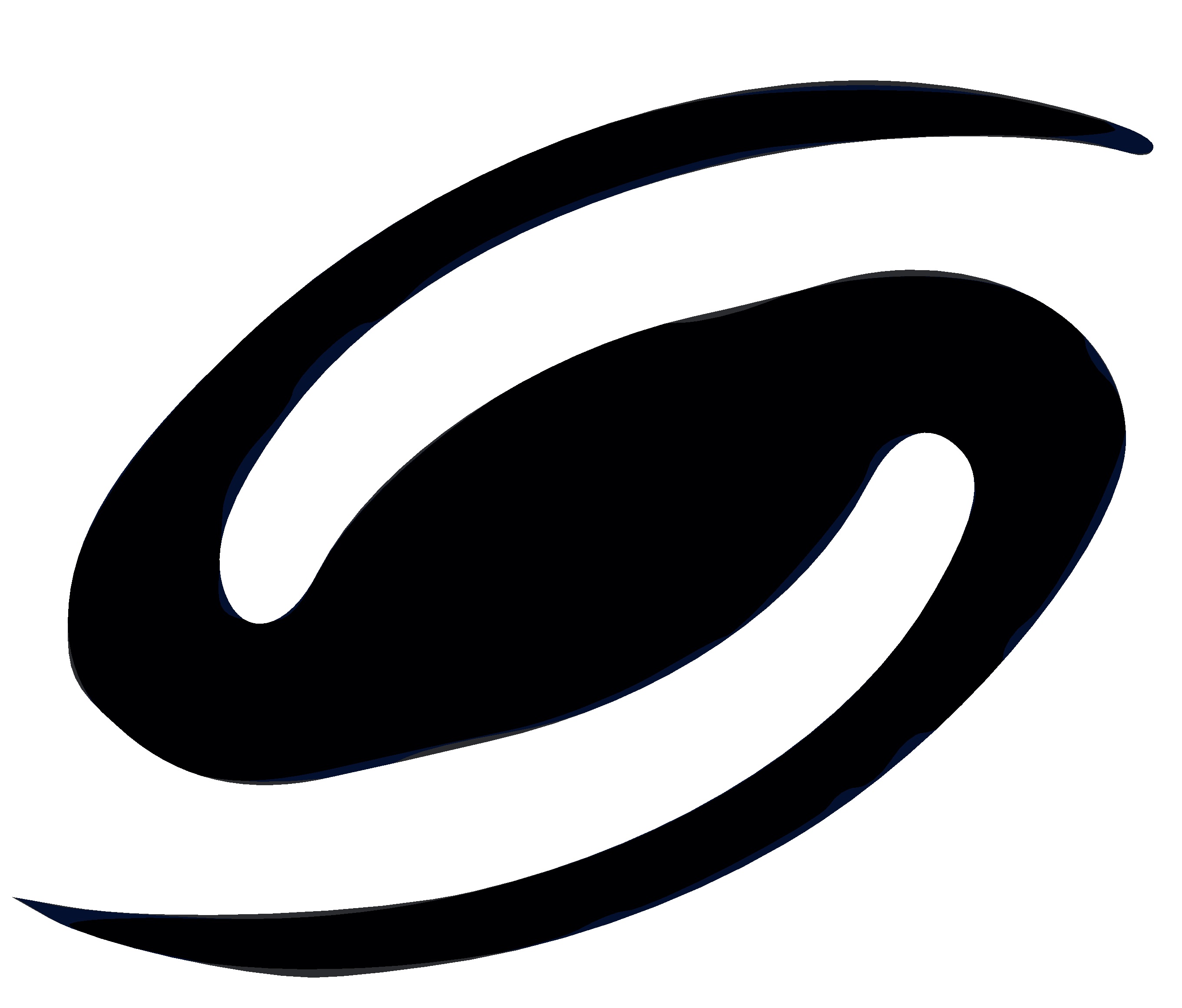}}
\usepackage{amstext}

\begin{document}

\title{Supermassive black hole wake or bulgeless edge-on galaxy?}
\subtitle{II: Order-of-magnitude analysis of the two physical scenarios} 

   \author{J. S\'anchez Almeida
          \inst{1,2}
          }

\institute{Instituto de Astrof\'\i sica de Canarias, c/ Vía Láctea s/n, E-38205, La Laguna, Tenerife,  Spain \\
        \email{jos@iac.es}
      \and
      Departamento de Astrof\'\i sica, Universidad de La Laguna, E-38203, La Laguna, Tenerife, Spain
      }

   \date{Received \today; accepted \dots }

 
   \abstract
{A recently discovered thin long object aligned with a nearby galaxy could be the stellar wake induced by the passage of a supermassive black hole (SMBH) kicked out from the nearby galaxy by the slingshot effect of a three-body encounter of  SMBHs. Alternatively, the object could be a bulgeless edge-on galaxy coincidentally aligned with a second nearby companion.  In  contrast with the latter, the SMBH interpretation requires a number of unlikely events to happen simultaneously.}
{We aim to assign a probability of occurrence to the  two competing scenarios.}
{The probability that the SMBH passage leaves a trace of stars is factorized as the product of the probabilities of all the independent events required for this to happen ($P_{\rm SMBH}$). Then, each factor is estimated individually. The same exercise is repeated with the edge-on galaxy interpretation ($P_{galax}$).}
{Our estimate yields $\log(P_{galax}/P_{\rm SMBH}) \simeq 11.4\pm 1.6$, where the error is evaluated considering that both $P_{galax}$ and $P_{\rm SMBH}$ are products of a large number of random independent variables. Based on the estimated probabilities, $P_{\rm SMBH} < 6\times 10^{-17}$ and $P_{galax} > 1.4\times 10^{-5}$, we determined the number of objects to be expected in various existing, ongoing, and forthcoming surveys, as well as among all observable galaxies (i.e., when observing between $10^6$ and $2\times 10^{12}$ galaxies). In the edge-on galaxy scenario, there are always objects to be detected, whereas  in the SMBH scenario, the expectation is always compatible with zero.}
{Despite the appeal of the runaway SMBH explanation, arguments based on the Occam's razor clearly favor the bulgeless edge-on galaxy interpretation.
  Our work does not rule out the existence of runaway SMBHs leaving stellar trails. It tells that the vD23 object is more likely to be a bulgeless edge-on galaxy.
}
   \keywords{Galaxies: halos ---
  Galaxies: kinematics and dynamics ---
   Galaxies: fundamental parameters ---
   Galaxies: peculiar --- 
  Galaxies: structure }

\titlerunning{Supermassive black hole wake or bulgeless edge-on galaxy? II}
   \maketitle
   
%

   \section{Introduction}\label{sec:intro}
\begin{figure}
     \includegraphics[width=9cm]{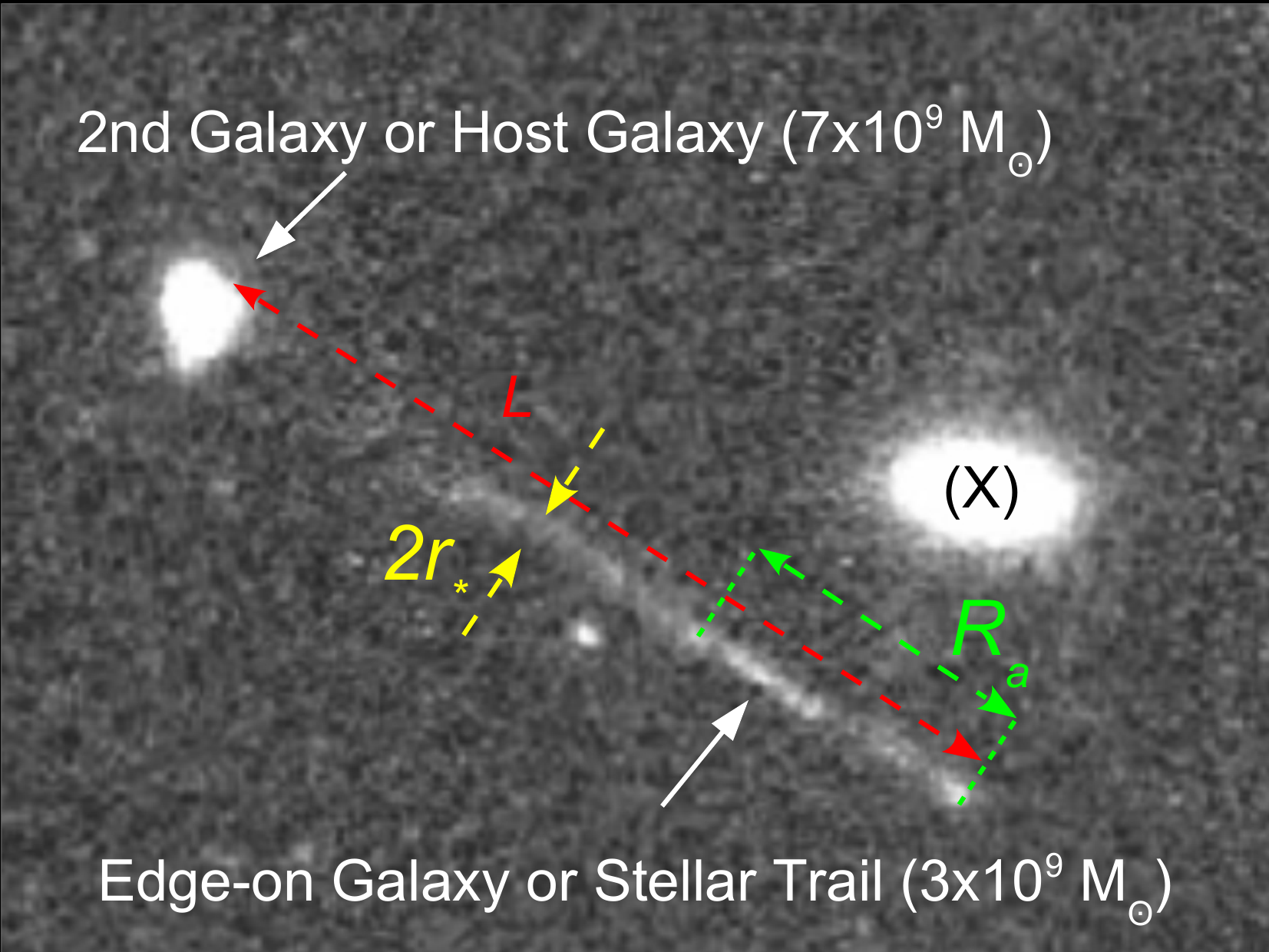}
     \caption{Graphical definition of  the main observational parameters characterizing the long thin stellar structure discovered by vD23. We analyze two interpretations: a stellar wake produced by a runaway SMBH escaping from the host galaxy, and a disk of a bulgeless edge-on galaxy randomly associated with a second galaxy. The image, taken from vD23, corresponds to the rest-frame UV because the objects are at redshift $\sim 1$ and were observed at wavelengths between 600 and 800\,nm ($L\simeq 60\,{\rm kpc}$, $R_a\simeq 25\,{\rm kpc}$, and $r_\star\simeq 1\,{\rm kpc} $; see Table~\ref{tab:parameters}). The corresponding stellar masses are given in parentheses. The galaxy marked with a cross is seen in projection and does not belong to the analyzed scene.}
  \label{fig:parameters}
\end{figure}

Figure~\ref{fig:parameters} shows a thin long object  ($\sim$0\farcs 25\,$\times$\,5\arcsec) aligned with a nearby galaxy discovered by \citet[][hereafter vD23]{2023ApJ...946L..50V}. It was interpreted as the stellar wake induced by the passage of a supermassive black hole (SMBH) that was kicked out from the nearby galaxy by the slingshot effect of a three-body encounter of SMBHs.
If this eye-catching interpretation were correct, it provides a unique way of detecting runaway SMBHs and giant molecular clouds in the circumgalactic medium (CGM) of galaxies. However, the interpretation  requires a number of unlikely events to happen simultaneously: (1) the three-body encounter of SMBHs, (2) the existence of a giant gas cloud in the CGM aligned with the direction in which the SMBH was ejected, (3) the gas cloud to be gravitationally unstable and so ready to collapse into stars, and (4) the newly formed stars to be on the Tully-Fisher relation (TFR), characteristic of galaxies.  The last requirement involves some kind of connection between the mass in newly formed stars and the proper motions of the original gas cloud, which have to be fine-tuned for the resulting  object to be on the TFR.

The likelihood of all these events to happen is relevant in the context of the second more canonical explanation we put forward in \citet[][hereafter Paper~I]{2023A&A...673L...9A}. The thin stellar structure shown in Fig.~\ref{fig:parameters} could be a long bulgeless galaxy viewed edge-on. The knotty appearance of the edge-on disk would arise because the image corresponds to the rest-frame UV.  In this alternative explanation, the only unusual condition to be requested is the coincidental alignment between the edge-on galaxy and a nearby companion. Bulgeless edge-on galaxies are already known to exist: Paper~I gives the example of IC\,5249, but many others could be listed \citep[e.g.,][]{1999BSAO...47....5K}.  Because the object is a disk galaxy, the stellar trace is automatically on the TFR without any fine-tuning.

Thus, despite the appeal of the runaway SMBH explanation, Occam's razor arguments  naturally favor the bulgeless edge-on galaxy interpretation. In this follow-up of Paper~I,  we quantify this claim by estimating the probability of the different physical processes and the conditions needed for the two scenarios to hold.
The formation of a stellar trail from the passage of an SMBH is examined in Sect.~\ref{sec:smbh_scenario}.
  We assume the need for three-SMBH encounters, according to the main scenario put forward by vD23. However, the recoil induced by the emission of gravitational waves during a binary SMBH merger \citep[e.g.,][]{2007PhRvL..98w1102C,2012PhRvD..85h4015L} can also produce the ejection of the final SMBH. This alternative channel is also mentioned in vD23, and Appendix~\ref{app:twosmbhs} considers it. It leads to slightly larger probabilities without altering the overall picture or the conclusions presented in Sect.~\ref{sec:smbh_scenario}.
The bulgeless edge-on scenario is studied in Sect.~\ref{sec:alignment}.
The probabilities resulting from these analyses are summarized in Table~\ref{tab:summary}.
The error bars of individual probabilities are extremely uncertain, but the central limit theorem can be invoked to estimate the error on the ratio of probabilities corresponding to the two scenarios. This exercise is carried out  in Sect.~\ref{sec:conclusions}, where  the conclusions of the analysis are considered. The discussion includes determining the number of stellar trails to be expected in various existing, ongoing, and forthcoming galaxy surveys, as well as the number of trails expected among all galaxies in the observable Universe.
In the case of the SMBH scenario, the expected number of trails in all cases is compatible with zero. 
Throughout the paper, the term {\em the object} refers to the long thin stellar trail shown in Fig.~\ref{fig:parameters}.


\section{Physical processes needed for the SMBH runaway scenario to work}\label{sec:smbh_scenario}

The probability of the SMBH leaving a trace of stars can be factorized as the product of all the independent events required for this to happen\footnote{This is very much in line with the famous Drake equation to estimate the number of alien civilizations in the Milky Way that are able to actively communicate with us \citep{1961PhT....14d..40D}.}, namely,
\begin{equation}
  P_{\rm SMBH}=P_{\rm 3\bullet}\,P_{eject}\,P_{fila}\,P_{shot}\,P_{unst}\,P_{\rm TFR1},
  \label{eq:psmbh}
\end{equation}
where the different symbols stand for:
$P_{\rm 3\bullet}$ , the probability that a randomly chosen galaxy has three nuclear SMBH orbiting each other,
$P_{eject}$ , the probability that the three SMBHs are ejected from the galaxy,
$P_{fila}$ , the probability that the galaxy CGM has a massive radial gas filament,
$P_{shot}$ , the probability that one of the SMBHs is shot right along the axis of the existing filament,
$P_{unst}$ , the probability that the passage of the SMBH triggers star formation in the filament,
and $P_{\rm TFR1}$ , the probability that the formed stellar mass is on the TFR. 
The different probabilities are estimated in the forthcoming sections, and Table~\ref{tab:summary} summarizes their values and the relevant equations. Based on these values,
\begin{equation}
P_{\rm SMBH} < 6\times 10^{-17},
  \label{eq:psmbh2}
\end{equation}
a truly small number compared with the number of known galaxies and even the number of galaxies expected in the visible Universe. The implications of the smallness of $P_{\rm SMBH}$   are discussed in detail in Sect.~\ref{sec:conclusions}.

  As we mentioned in Sect.~\ref{sec:intro}, the three-SMBH encounter is the main scenario supported by vD23, but they also offered another channel to produce runaway SMBHs, namely, the recoil induced by the emission of gravitational waves during a binary SMBH merger.  This alternative channel is also explored in Appendix~\ref{app:twosmbhs}, and it leads to a probability  that is about 20 times larger than that in Eq.~(\ref{eq:psmbh2}) (Eq.~[\ref{eq:psmbh2_2}]). This increase, however, does not modify the conclusions of this work (Sect.~\ref{sec:conclusions}).

\begin{table*}
\caption{Evaluated probabilities}              
\label{tab:summary}      
\centering                                      
\begin{tabular}{l c l l}          
\hline\hline                        
Symbol& Value & Equation & Comments\\    
\hline                                   
  $P_{{\rm 3\bullet}}$ &$<2.5\times 10^{-2}$ & ~~~~(\ref{eq:pof3})  & Probability that a galaxy has 3 SMBHs orbiting each other at any time\\
  $P_{eject}$  & $10^{-3}$ &~~~~(\ref{eq:peject}) & Probability that the 3 SMBHs are ejected from the galaxy \\
  $P_{fila}$&$2.2\times 10^{-2}$ &~~~~(\ref{eq:pfila})& Probability that the galaxy has a massive radial gas filament\\
  $P_{shot}$&$6.9\times 10^{-5}$&~~~~(\ref{eq:pshot})& Probability that the SMBH is shot right along the axis of the existing filament\\
  $P_{unst}$&$<3.0\times 10^{-5}$&~~~\,(\ref{eq:punst})& Probability that the passage of the SMBH triggers star formation in the filament\\
  $P_{{\rm TFR1}}$&$<5.7\times 10^{-2}$&~~~\,(\ref{eq:ptf1})& Probability that the formed stars are on the TFR\\
  $P_{\rm SMBH}$ &$ < 6\times 10^{-17}$&~~(\ref{eq:psmbh})(\ref{eq:psmbh2}) & Overall probability of the runaway SMBH scenario\\
--- &&&~~~~~~~~~~~~~---\\
$P_{\galaxy bl}$&$\gtrsim 0.2$&~~~\,(\ref{eq:pbl})& Probability for a  galaxy to be bulgeless\\
  $P_{\galaxy eo}$ &$8\times 10^{-2}$&~~~\,(\ref{eq:pgeo})& Probability for a disk galaxy to be observed edge-on\\

$P_D$  &$\gtrsim 3\times 10^{-2}$&~~~\,(\ref{eq:pd})&Probability of having a second galaxy close enough\\
$P_{align}$ &$2.9\times 10^{-2}$&~~~\,(\ref{eq:p2nd})& Probability of having the second galaxy along the major axis of the projected disk\\
  $P_{{\rm TFR2}}$&1&~~~\,(\ref{eq:ptf2})& Probability that the formed stars are on the TFR\\
  $ P_{\rm galax}$ &$\gtrsim 1.4\times 10^{-5}$&~(\ref{eq:pgalax})(\ref{eq:pgalax2})& Overall probability of the edge-on galaxy scenario\\
--- &&&~~~~~~~~~~~~~---\\
  $N_s$  &$\sim 2\times 10^8$&& Number of galaxies in the largest existing surveys$^a$\\
$N_s P_{\rm SMBH}$   &$1.2\times 10^{-8}$&& Expected number of objects in the SMBH scenario\\
$N_s P_{galax}$   &$1.4\times 10^{4}$&& Expected number of objects in the edge-on galaxy scenario\\
--- &&&~~~~~~~~~~~~~---\\
  $N_U$  &$\sim 2\times 10^{12}$&& Expected number of galaxies in the visible Universe$^b$\\
$N_U P_{\rm SMBH}$   &$1.2\times 10^{-4}$&& Expected number of objects in the SMBH scenario\\
$N_U P_{galax}$   &$1.4\times 10^{8}$&& Expected number of objects in the edge-on galaxy scenario\\
\hline                                             
\end{tabular}
\begin{flushleft}
  $^a$~For example, the Dark Energy Survey described by \citet[][]{2021ApJS..255...20A}.

  $^b$~Up to redshift eight, according to  \citet[][]{2016ApJ...830...83C}. 
\end{flushleft}
\end{table*}
\begin{table*}
\caption{Parameters characterizing the scenarios}              
\label{tab:parameters}      
\centering                                      
\begin{tabular}{lccl}          
\hline\hline                        
~~~~Parameter& Symbol$^a$ & Value & Comments\\    
  \hline                                   
--- General ---&&&\\
  Distance to the tip of the object&$L$&60\,kpc&Fig.~\ref{fig:parameters}\\
  Width of the object&$2r_\star$& 2\,kpc&Fig.~\ref{fig:parameters}\\
 Stellar mass of the 2nd galaxy&&$7\times 10^{9}\,{\rm M_\odot}$&Fig.~\ref{fig:parameters}; vD23\\
  Stellar mass of the object&&$3\times 10^{9}\,{\rm M_\odot}$&Fig.~\ref{fig:parameters}; vD23\\
  Redshift of the object&&1&vD23\\
  Tolerance in the TFR&$\Delta \log V_{max}$& 0.1\,dex& S\'anchez~Almeida et al. (2023) \\ 
  Present age of the Universe&$t_H$&13.5\,Gyr&\\
--- Runaway SMBH ---&&&\\
  Star formation efficiency &&${\rm 1\,Gyr^{-1}}$&\citet[][]{2008AJ....136.2782L}\\
  Age of the stellar trail&&39\,Myr&vD23\\
  Gas mass needed to form the object&$M_g$&${\rm 7.7\times 10^{10}\,M_\odot}$& Sect.~\ref{app:cgmgas}\\
  Sound speed &$c_s$&${\rm 10\,km\,s^{-1}}$&Sect.~\ref{app:cgmgas}\\
  Temperature of the gas&&$10^4\,{\rm K}$&Sect.~\ref{app:cgmgas}\\
  Jeans length&$\lambda_J$&$5\times 10^{-2}$\,kpc&Sect.~\ref{app:cgmgas}\\
  Free fall time in the halo of the host&$t_{fila}$&150\,Myr&Sect.~\ref{app:cgmgas}\\
  Number of big filaments per galaxy&$n_{fila}$&2&Sect.~\ref{app:cgmgas}\\
  Mass of the SMBH&$M_\bullet$&${\rm 7\times 10^{6}\,M_\odot}$&Sect.~\ref{app:appa}; vD23\\
  Velocity of the SMBH& $v$& 1600\,km\,s$^{-1}$& vD23\\
  Width of the wake&$2d$&2\,kpc& Fig.~\ref{fig:wake}, with $d=r_\star$\\
  Distance SMBH to gas clouds &$d_{min}$&1\,kpc&$d_{min}=r_\star$;~Eq.~(\ref{eq:deltau})\\
  Density of the gas &$\rho$&$7\times 10^{-25}\,{\rm g\,cm^{-3}}$&Sects.~\ref{app:cgmgas} and \ref{app:appa}\\
  Bondi–Hoyle–Lyttleton accretion radius & $R_{acc}$ & $2.3\times 10^{-2}\,{\rm pc}$& Eq.~(\ref{eq:racc})\\
  Galaxies without stellar trail  &$N_{no}$&$1.5\times 10^6$&SDSS, \citet[][]{2012ApJS..203...21A}\\
  Thermal velocity of the gas&$v_{ther}$& ${\rm \sim 7\,km\,s^{-1}}$& Sect.~\ref{sec:waketfr}\\
  Scape velocity from the 2nd galaxy&$v_{esc}$& ${\rm \sim 400\,km\,s^{-1}}$& Sect.~\ref{sec:waketfr}\\
  --- Bulgeless edge-on galaxy ---&&&\\
Radius of the edge-on disk&$R_a$&25\,kpc&Figs.~\ref{fig:parameters} and \ref{fig:edge-on-schematic}\\
  Full width of the edge-on disk&$W_a$&2\,kpc&$W_a=2r_\star$; Sect.~\ref{sec:alignment}\\
  Number surface density of galaxies& $\Sigma_g$&${\rm > 1\,Mpc^{-2}}$&Sect.~\ref{sec:alignment}\\
    Full width of the 2nd galaxy&$W_b$&6\,kpc&$W_b=3W_a$; Fig.~\ref{fig:edge-on-schematic}b\\
\hline                                             
\end{tabular}
  \begin{flushleft}
  $^a$~Empty boxes correspond to parameters without specific symbols in the main text. 
\end{flushleft}
\end{table*}

%

\subsection{The three-body SMBH encounter}\label{sec:three_body}

The process of kicking out  three SMBHs from the center of a galaxy is very complex and far from being properly understood and modeled. It requires the formation of a triple system from two successive major mergers that occur with a time lag short enough to prevent the coalescence of the first SMBH with the SMBH already existing at the center of the galaxy.  When a triple system is formed, it has to evolve to kick out the three bodies from the galaxy, as required in the scenario proposed by vD23. The three bodies escape to conserve linear momentum, provided the mass of the three SMBHs is not very different,  and keeping in mind the large distance from the central galaxy of the SMBH that created the stellar trail in this scenario  ($L\sim 60~{\rm kpc}$;  Fig.~\ref{fig:parameters} and Table~\ref{tab:parameters}).
For the SMBHs to escape from the host galaxy, the three-body interaction has to produce an extreme slingshot effect.        
\citet{2007MNRAS.377..957H} followed the evolution of a triple SMBH system after the system is formed in the center of a massive galaxy. Their modeling includes the dynamical friction produced by stars and dark matter, as well as the emission of gravitational waves. They considered a large number of initial conditions appropriate for early-type galaxies.  For this particular modeling, they concluded that the escape of all three SMBH is exceedingly rare, about 0.1\,\% of the $\sim 8000$ simulated systems. Thus, we take
\begin{equation}
  P_{eject}\simeq 10^{-3}.
\label{eq:peject}
\end{equation}

The fine-tuning required to form a triple SMBH system (two consecutive major mergers in which the SMBHs do not merge in pairs, but form a triple system) suggests that most galaxies will not go through the formation of one such system during their lifetimes. The fact that three galaxies are needed to form one system implies that fewer than one-third of the galaxies would undergo this phase.  Even if the lifetime of the triple system could be long \citep[1\,Gyr;][]{2007MNRAS.377..957H}, it has to be present at around the moment when the galaxy was observed and the SMBH ejected. Thus, the probability for a galaxy to have one of them at any given time would be   
\begin{equation}
  P_{\rm 3\bullet} < \frac{1\,{\rm Gyr}}{3\,t_H}\simeq 2.5\times 10^{-2},
  \label{eq:pof3}
\end{equation}
with $t_H$ the age of the Universe, taken to be 13.5\,Gyr.  

\subsection{Conditions for stars to form in the CGM}\label{app:cgmgas}

The CGM of galaxies is expected to be rich in gas, filled with gas accreted from the intergalactic medium (IGM), debris from satellites,  and ejecta from star formation and active galactic nucleus (AGN) activity. This CGM gas has multiple phases with different temperatures, densities, chemical compositions, and kinematics \citep[e.g.,][]{2014A&ARv..22...71S}. In halos larger than $10^{12}\,{\rm M_\odot}$, most of this gas is at virial temperatures of millions of K because the gas accreted from the IGM is shock-heated when encountering preexisting CGM gas \citep[e.g.,][]{1977MNRAS.179..541R,1978MNRAS.183..341W,2003MNRAS.345..349B}. Below this halo-mass threshold, the gas may be colder since it is partly accreted directly from the IGM in the so-called cold-flow  mode \citep[][]{2009Natur.457..451D}, but its characteristic temperature still exceeds $10^4\,{\rm K}$ \citep[e.g.,][]{2012MNRAS.423.2991V}. The gas from a cosmological origin is expected to be metal poor, and it coexists with metal-rich gas dragged along by stellar winds, supernova (SN)-driven bubbles, and jets from the central galaxy \citep[e.g.,][]{2014IAUS..298..228F}.  None of these gas phases is expected to be Jeans unstable, and it is therefore not ready to form stars upon small perturbations. This picture of the CGM physical conditions does not prevent the existence of pockets of gas that could be cold, dense, and unstable, but the bulk of the CGM gas will not be in such a state. In other words, it is not straightforward to produce the observed $3\times 10^9\,{\rm M_\odot}$ in stars (Table~\ref{tab:parameters}) out of the gas expected in the CGM.

To evidence the magnitude of the difficulty, consider  a realistic star-formation efficiency of $1\,{\rm Gyr}^{-1}$ \citep[e.g.,][]{2008AJ....136.2782L}.  To form $3\times 10^{9}\,{\rm M_\odot}$ in 39\,Myr, we need a gas mass of $M_g \sim 7.7\times 10^{10}\,{\rm M_\odot}$ ($\simeq 3\times 10^{9}\,{\rm M_\odot \times 1\,Gyr/ 39\,Myr}$).
The host galaxy in which the SMBHs originally resided has a stellar mass $M_\star \simeq 7\times 10^{9}\,{\rm M_\odot}$ (Table~\ref{tab:summary}), which corresponds to a dark matter halo mass of  $M_h\sim 3\times 10^{11}\,{\rm M_\odot}$ \citep[e.g.,][Figs. 11 and 13]{2010ApJ...717..379B}. Thus the host galaxy should have a baryon fraction $(M_\star+M_g)/(M_h+M_\star+M_g)$ $\simeq 0.22$, which is much larger than the cosmic baryon fraction of 0.16 \citep[][]{2016A&A...594A..13P}. This difficulty stems from the enormous gas mass needed to form  $3\times 10^{9}\,{\rm M_\odot}$ of stars in only  39\,Myr.

A simple order-of-magnitude estimate shows that the runaway SMBH scenario requires all the employed gas to be concentrated along the direction where the SMBH was shoot.  The reasoning is that  the protocloud from which the stars originate cannot be much larger than the observed stellar trail because the Jeans length $\lambda_J$, characterizing the size of the region that collapsed to form stars, must be quite small,
\begin{equation}
\lambda_J=c_s\,t_{f\!\!f}\simeq 0.05\,{\rm kpc},
\end{equation}
with $t_{f\!\!f}$ the characteristic timescale for the gravitation instability to proceed (i.e., the free-fall timescale of the self-gravitating gas clouds), and $c_s$ the sound speed \citep[e.g.,][]{2008gady.book.....B}. The numerical value for $\lambda_J$ comes from  $t_{f\!\!f}\lesssim 5\,{\rm Myr}$, which is the timescale required to form the youngest stars of the trail (the oldest have 39\,Myr), and $c_s\simeq 10\,{\rm km\,s^{-1}}$, which is set by the gas temperature, assumed to be $10^4\,{\rm K}$, as expected for the cold component of the CGM \citep[e.g.,][]{2012MNRAS.423.2991V}. The Jeans length is clearly within the lateral size of the trail, a conclusion that holds independently of the uncertainties in $t_{f\!\!f}$ and $c_s$. In short, the required protocloud of  $\sim 10^{11}\,{\rm M_\odot}$ must have been close in location and in shape to the observed stellar trace.
Another independent argument also points out that the direction in which the SMBH was ejected must have been very special. If this were not the case and the CGM around the host galaxy were rather isotropic, it would have a gas mass  of $\sim 5\times
10^{14}\,{\rm M_\odot}$, that is, the total mass of a massive galaxy cluster because the volume occupied by the observed stellar trail ($\sim\pi\times [1\,{\rm kpc}]^2\times 60\,{\rm kpc}$) is only about $2\times 10^{-4}$ times the volume of the CGM around the central galaxy ($\sim [4\pi/3]\times 60^3 \,{\rm kpc}^{3}$). 

Thus, the direction of the SMBH ejection must have been extremely well fine-tuned to hit a preexisting filamentary gas cloud. To estimate the probability of this happening, we begin by acknowledging that the alignment of CGM gas clouds in filaments may happen in a scenario in which the gas is accreted through cosmological cold-flows \citep[e.g.,][]{2009Natur.457..451D}. The chances that one such filament contains all the gas in the halo at the moment of the SMBH ejection are unknown, but should be about
\begin{equation}
  P_{fila}=n_{fila}\,t_{fila}/t_H\simeq 2.2\times 10^{-2},
  \label{eq:pfila}
\end{equation}
with $t_{fila}$ the lifetime of one such filament,  $n_{fila}$ the number of times in the age of the Universe in which one of these major cosmological gas accretion events happens, and $t_H$ the age of the Universe. The lifetime of one of these filaments is short, about the free-fall time in a halo of the galaxy \citep[e.g.,][]{2014A&ARv..22...71S}. This time only depends on the average density of the halo, which for DM halos is about $10^4$ times the average density of the Universe \citep[e.g.,][]{2015MNRAS.452.1217C}. At $z=1, \text{this}$ renders $t_{fila}\simeq 150\,{\rm Myr}$. Using $n_{fila}=2$ and $t_H=13.5\,{\rm Gyr}$, we find the numerical value included in Eq.~(\ref{eq:pfila}).

When the original gas filament is in place, the runaway SMBH has to pierce it  along its major axis. Because it could have been shot in any direction, the probability of hitting the target is
\begin{equation}
  P_{shot} =  \frac{\pi (r_\star/L)^2}{4\pi}\simeq 6.9\times 10^{-5},
  \label{eq:pshot}
\end{equation}
with $\pi (r_\star/L)^2$ the solid angle of the outermost extreme of the stellar trail as seen from the host galaxy, and $4\pi$ the solid angle of the whole sky (Fig.~\ref{fig:parameters}). The numerical value in Eq.~(\ref{eq:pshot}) uses $r_\star=1\,{\rm kpc}$ and $L = 60\,{\rm  kpc}$.

\subsection{Perturbations expected from the passage of a fast-moving SMBH through a material medium }\label{app:appa}

In this section, we examine various physical mechanisms by which the passage of an SMBH disturbs the surrounding medium. We will show that  given the conditions required by the vD23 scenario (high speeds and relatively large spatial scales), the expected impact on the medium is negligibly small. Therefore, for a gas cloud to collapse by the passage of an SMBH, it has to be already Jeans unstable. In this case, the probability of an SMBH passage to trigger star formation becomes the probability for a galaxy to have a large massive Jean unstable gas filament at the particular time when the SMBH is shot. This type of unstable filaments does not seem to be present in observed galaxies, a fact that we use to set an upper limit to the probability of triggering star formation by the passage of the SMBH.    

\begin{figure}
   \centering
   \includegraphics[width=9cm]{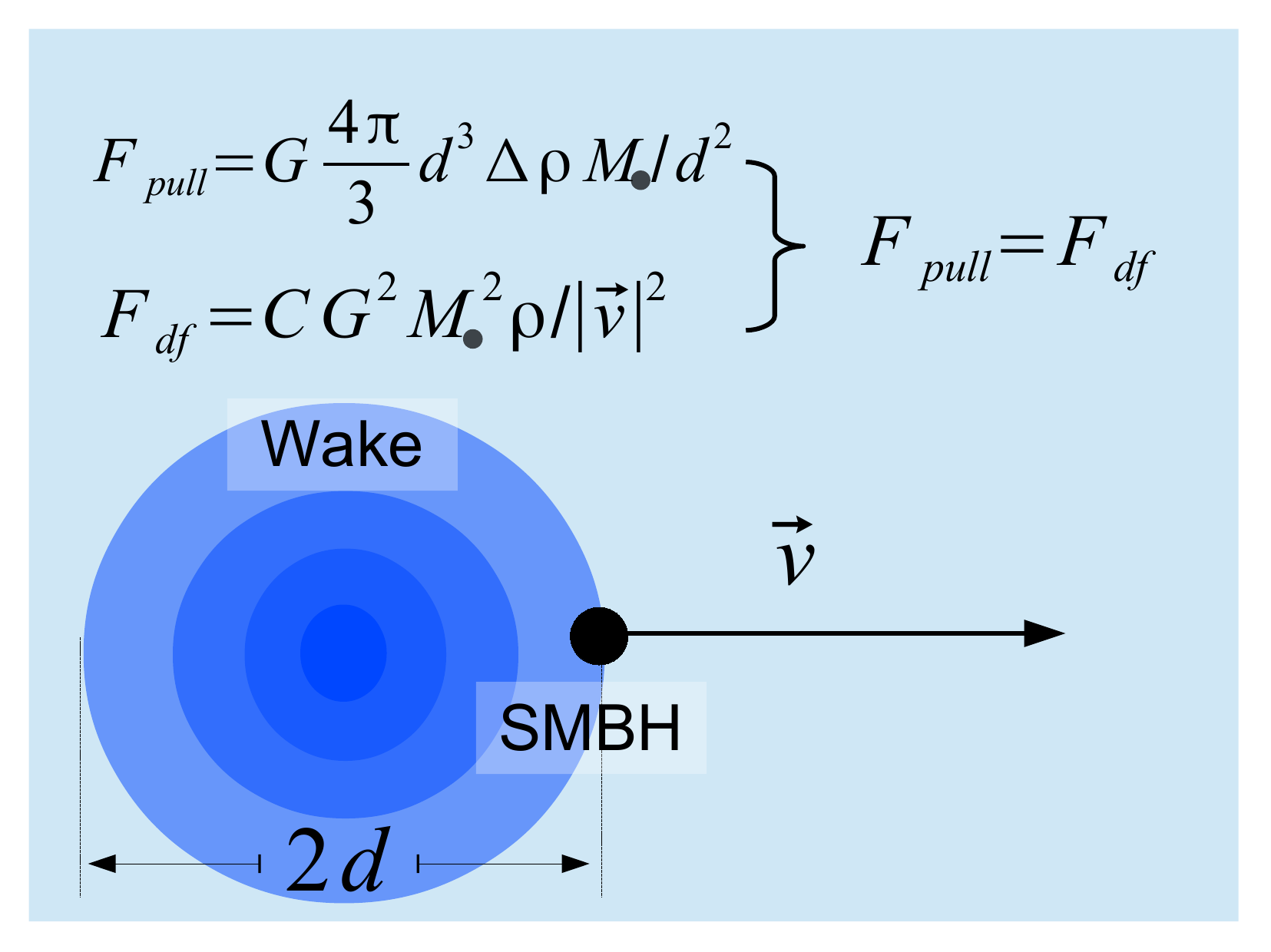}
   \caption{Schematic showing how the overdensity of the wake $\Delta\rho$ can be estimated by equating its gravitational pull ($F_{pull}$) with the dynamical friction force ($F_{df}$) when an SMBH moves with a velocity
     $\vec{v\,}$
     in a medium with a density $\rho$. $C$ in $F_{df}$ is just a scaling constant. See the main text for further details and references. 
   }
\label{fig:wake}%
\end{figure}
An SMBH moving through a background of collisionless particles induces an overdensity of the particles left behind, which, through its gravitational pull, causes \citeauthor{1943ApJ....97..255C}'s dynamical friction \citep[e.g.,][]{1971Ap&SS..13..279K,1983A&A...117....9M,2022MNRAS.515..407K}. The overdensity of the wake can easily be estimated by equating the dynamical friction force \citep[e.g.,][]{2008gady.book.....B} and the additional gravitational pull (see Fig.~\ref{fig:wake}). The exercise renders
\begin{equation}
\frac{\Delta\rho}{\rho}\simeq K \frac{G\,M_\bullet}{v^2\,d}
\simeq 1.2\times 10^{-5} \frac{(M_\bullet/7\times 10^6\,{\rm M}_\odot)}{(v/1600\,{\rm km\,s^{-1}})^{2}\,(d/1\,{\rm kpc})},
\label{eq:overdensity}
\end{equation}
where $K$ is a numerical value of order one, $G$ stands for the gravitational constant, $M_\bullet$ represents the SMBH mass, $v$ is the relative velocity, and $d$ corresponds to the radius (or to a characteristic length scale) of the induced overdensity. This approximate formula agrees well with the wake found through numerical simulations of DM subhalos moving in a medium of constant density \citep[e.g.,][]{2018PhRvL.120u1101B}\footnote{They modeled a DM subhalo of $2\times 10^7\,{\rm M}_\odot$ moving with a relative speed of $200\,{\rm km\,s^{-1}}$ and leaving a wake at a distance of approximately 0.5\,kpc. They find a peak overdensity $\lesssim 0.01,$ whereas  Eq.~(\ref{eq:overdensity}) predicts $\Delta\rho/\rho\simeq 0.004$.}.
The scenario invoked by vD23 has $M_\bullet\simeq 7\times 10^6\,{\rm M_\odot}$\footnote{One-third of the total black hole mass expected for the host galaxy ejecting the SMBH, provided it follows the stellar mass--black hole  mass relation by \citet{2019ApJ...887..245S} with $M_\star\sim$$7\times 10^9\,{\rm M_\odot}$.}, $v=1600\,{\rm km\,s^{-1}}$, and $d\sim 1\,{\rm kpc}$ (Table~\ref{tab:parameters}), so that the passage of the SMBH induces an overdensity of only $\Delta\rho/\rho \sim 10^{-5}$ . In this estimate, $d$ has been set rather arbitrarily to half the width of the object ($d=r_\star$, Fig.~\ref{fig:parameters}), but it can be changed by orders of magnitude, and the induced overdensity still remains negligibly low.

\citeauthor{1943ApJ....97..255C}'s drag is derived for an external medium made of collisionless particles, however, the drag is very similar when applied to an object moving with highly supersonic speed in a gaseous external medium \citep[see][their Fig.~3]{1999ApJ...513..252O}. Thus, Eq.~(\ref{eq:overdensity}) provides the correct order-of-magnitude estimate even for the motion through gaseous media. This holds true even in the model by \citet{1972ApL....11...87S}, which invokes the shock wave created by the SMBH in the gas. They determined the drag force on the SMBH created by this wake, which turns out to be a combination of \citeauthor{1943ApJ....97..255C}'s drag and the gas accretion cross-section of the massive object. As we show below, the effective cross section of an acreting SMBH is irrelevant in this context, and the formulae are reduced to Eq.~(\ref{eq:overdensity}).   

Another interaction mechanism was proposed by \citet{2008ApJ...677L..47D}, who determined the relative velocities $\Delta U$ induced on a nearby molecular cloud by the tidal forces of an SMBH passing by at a distance of $d_{min}$, 
\begin{equation}
  \Delta U \simeq 3.8\times 10^{-2}\,{\rm km\,s^{-1}} \frac{(M_\bullet/7\times 10^6\,{\rm M}_\odot)}{(v/1600\,{\rm km\,s^{-1}})\,(d_{min}/1\,{\rm kpc})}.
  \label{eq:deltau}
\end{equation}
For the conditions of the rogue SMBH proposed by vD23 (Table~\ref{tab:parameters}), the induced velocity perturbations, $\sim 0.04\,{\rm km\,s^{-1}}$, can hardly trigger the runaway process needed for the star formation to occur. For example,  thermal and turbulent velocities are much larger than $\Delta U$ and are therefore liable to trigger instabilities before tidally induced velocities can do it. 

A form of interaction between a traveling SMBH and the surrounding gaseous medium is through the accretion of all the mass within its cross section. There is no evidence of strong AGN activity in the scenario proposed by vD23, therefore, we can assume that this cross section is not set by the radius of a putative accretion disk surrounding the SMBH, but by the Bondi–Hoyle–Lyttleton accretion cross section, which is characterized by a radius $R_{acc}$ given by \citep[e.g.,][]{2004NewAR..48..843E}
\begin{equation}
  R_{acc}=\frac{2\,G\,M_{\bullet}}{v^2} = 2.4\times 10^{-5}\,{\rm kpc}\,  \frac{(M_\bullet/7\times 10^6\,{\rm M}_\odot)}{(v/1600\,{\rm km\,s^{-1}})^2},
  \label{eq:racc}
\end{equation}
which renders an accretion rate of
\begin{equation}
\frac{d M_{\bullet}}{dt}= \pi R_{acc}^2\,\rho\,v= \frac{4\pi G^2\,M_\bullet^2\,\rho}{v^3}, 
\end{equation}
and implies a total accreted gas mass during the passage of
\begin{equation}
  \frac{\Delta M_g}{M_g}\simeq \frac{dM_{\bullet}}{dt}\,\frac{L}{M_g\,v}\simeq
  \label{eq:accretedm}
\end{equation}
\begin{displaymath}
 ~~~1.1\times 10^{-11}\,\frac{(M_\bullet/7\times 10^6\,{\rm M}_\odot)^2}{(v/1600\,{\rm km\,s^{-1}})^4}\frac{L/60\,{\rm kpc}}{M_g/10^{11}\,{\rm M_\odot}}\,\frac{\rho}{7\times 10^{-25}\,{\rm g\,cm^{-3}}},
\end{displaymath}
with $L/v$ the crossing time of $\sim$39\,Myr. The density of the gas in Eq.~(\ref{eq:accretedm}) is equivalent to 0.3 atoms ${\rm cm^{-3}}$ and corresponds to the cold CGM at redshift one, as described in Sect.~\ref{app:cgmgas}.   The accreted gas mass is negligibly small.  Sound waves would restore the equilibrium filling the swept cylinder within a sound-wave crossing time, about a few thousand  years. This tiny perturbation results from the assumed SMBH cross section, but even an increase of this unknown by some orders of magnitude would leave the conclusion unchanged. 

In short, the various physical processes by which the passage of a fast-moving SMBH may potentially destabilize the surrounding medium induce negligible perturbations (Eqs.~[\ref{eq:overdensity}], [\ref{eq:deltau}], and [\ref{eq:accretedm}]), making it very unlikely that the passage alone can trigger significant star formation.  The perturbations are unable to compress a Jeans-stable gas to make it unstable. The preexisting gas filament should already be Jeans unstable so that any arbitrary small perturbation could trigger its  collapse to form stars. Thus, the probability of the SMBH triggering star formation is the probability of a galaxy having a Jeans-unstable gas filament when the SMBH passes by. This filament may eventually collapse into stars, regardless of whether an SMBH passes by. If these filaments were common, they would spontaneously produce stellar trails in the halos of the galaxies, which we do not observe\footnote{Faint slender stellar structures are sometimes observed, but they seem to be associated with tidal debris from old mergers or are triggered by an ongoing merger \citep[e.g.,][]{2005ApJ...631L..41M,2010AJ....140..962M,2021A&A...654A..40T,2023MNRAS.524.1431Z}.}.  This observational constraint can be used to set an upper limit on the probability of a filament to be Jeans unstable, $P_{unst}$. When in a set of $N_{no}$ randomly chosen galaxies only one shows a stellar trail (i.e., the object in the vD23 scenario), the probability for one of these galaxies to have a Jeans-unstable filament at present is  $N_{no}^{-1}$. This probability can be split into the probability of having a radial filament ($P_{fila}$ in Eq.~[\ref{eq:pfila}]) times $P_{unst}$, so that
\begin{equation}
  P_{unst} = \frac{1}{P_{fila}\,N_{no}} \leq 3.0\times 10^{-5}.
  \label{eq:punst}
\end{equation}
The numerical value in Eq.~(\ref{eq:punst}) uses $P_{fila}$ in Eq.~(\ref{eq:pfila}) and $N_{no} \geq 1.5\times 10^6$. This value for $N_{no}$ is the number of galaxies with images and spectra in the Sloan Digital Sky Survey \citep[SDSS; e.g.,][]{2011AJ....142...72E,2012ApJS..203...21A}, and in doing so, we assume that SDSS is deep enough to show stellar trails if they
  existed\footnote{The SDSS is almost complete for galaxies with $r$-band surface brightness lower than 24 mag\,arcsec$^{-2}$ \citep{2005ApJ...631..208B}. This cutoff is well above the expected surface brightness for the object in the optical (Paper~I).} and that none of the cataloged galaxies presents a conspicuous stellar trail similar to that in Fig.~\ref{fig:parameters}. This approach is probably quite conservative because ongoing galaxy surveys are orders of magnitude larger than the SDSS \citep[e.g., the Dark Energy Survey has $5\times 10^{8}$ galaxies;][]{2021ApJS..255...20A}, and no other stellar trail like the one in the vD23 scenario has been claimed so far. The main uncertainty in $N_{no}$ stems from the fact that a systematic search for companion stellar trails looking like the object has not been carried out.

\subsection{Stellar wake on the Tully-Fisher relation}\label{sec:waketfr}
The probability for a stellar wake to be on the TFR is one of the key unknowns of the probability in Eq.~(\ref{eq:psmbh}). In the SMBH scenario, the velocities in the position-velocity curve are random proper motions of the preexisting gas and  are therefore causally disconnected from the stellar mass formed by the passage of the SMBH. For lack of a better hypothesis, we assume that given the stellar mass of the stellar trail (Table~\ref{tab:parameters}), the maximum velocity ($V_{max}$) in the position-velocity curve that places the object on the TF relation \citep[Fig.~2 in][]{2023A&A...673L...9A} is a random variable. Specifically,  $\log V_{max}$ is a random variable with a uniform distribution that could be anywhere from the typical thermal velocities of the gas ($\log V_{ther}$) to the escape velocity within the host galaxy halo ($\log V_{esc}$),
\begin{equation}
  P_{\rm TFR1} < \frac{2\Delta \log V_{max}}{\log V_{esc}- \log V_{ther}}\simeq 5.7\times 10^{-2},
  \label{eq:ptf1}
\end{equation}
with $\Delta \log V_{max}$ the allowed separation from the TFR. The numerical value in Eq.~(\ref{eq:ptf1}) results from the following reasonable parameters:  $V_{ther}{\rm \sim 7\,km\,s^{-1}}$ (for a temperature of ${\rm 10^4\,K}$), $V_{esc}{\rm \sim 500\,km\,s^{-1}}$ \citep[for  $M_\star\sim  7\times 10^{9}\,{\rm M_{\odot}}$; e.g.,][]{2018ApJ...859..109S}, and $\Delta\log V_{max}\sim 0.05\,{\rm dex}$ \citep[][]{2023A&A...673L...9A}.
Equation~(\ref{eq:ptf1}) does not consider the fact that the position-velocity curve looks like the rotation curve of a galaxy (see Fig.~1e in Paper~I). Including this constraint would make $P_{\rm TFR1}$ much smaller, which is the reason why  Eq.~(\ref{eq:ptf1}) only sets an upper limit.

%
\section{Probability of a coincidental alignment between two galaxies}\label{sec:alignment}
%
\begin{figure}
     \includegraphics[width=9cm]{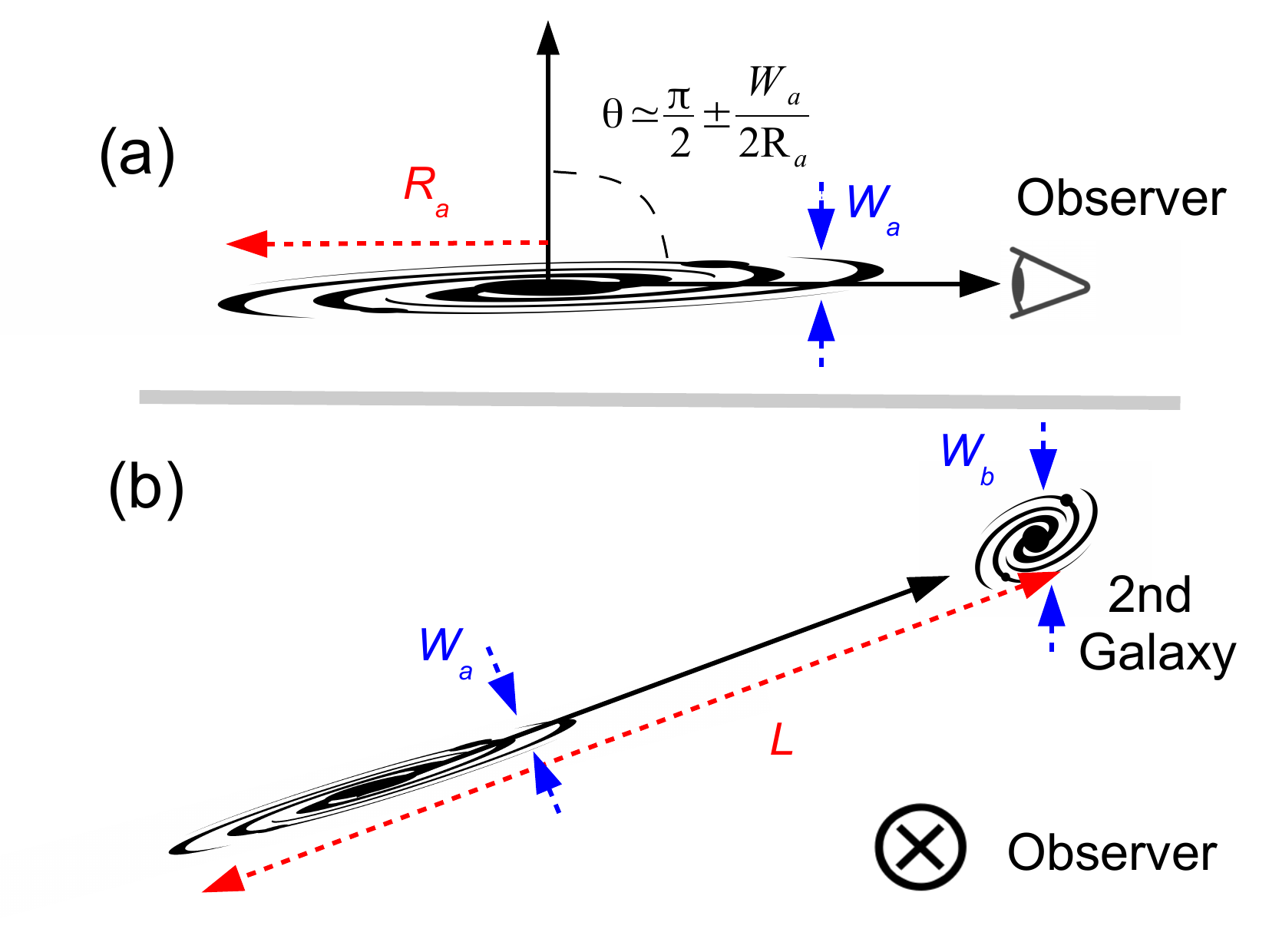}
     \caption{Diagrams to illustrate the edge-on-galaxy scenario.
       (a) Schematic with a side view of an almost edge-on galaxy. The angle between the line of sight and the normal to the disk, $\theta$, is almost $\pi/2$ in this orientation. The symbols $R_a$ and $W_a$ stand for the radius and the full width of the disk, respectively. (b) Schematic with a line-of-sight view of the edge-on galaxy and the second companion. The lengths $L$ and $R_a$ are the same as in Fig.~\ref{fig:parameters}, and $W_a \simeq 2r_\star$.}
 \label{fig:edge-on-schematic} 
\end{figure}

In this section, we analyze the second alternative scenario (see Fig.~\ref{fig:parameters}). The probability of finding a bulgeless edge-on galaxy
aligned with a second galaxy can be expressed as
\begin{equation}
  P_{galax} = P_{\galaxy bl}\,P_{\galaxy eo}\,P_D\,P_{align}\,P_{\rm TFR2},
  \label{eq:pgalax}
\end{equation}
with
$P_{\galaxy bl}$ the probability of galaxies to be bulgeless,
$P_{\galaxy eo}$ the probability of a disk galaxy to be observed edge-on,
$P_D$ the probability of having a second galaxy close enough, 
$P_{align}$ the probability of having this second galaxy along the major axis of the projected disk, and
$P_{\rm TFR2}$ the probability that the galaxy is on the TFR. We start from the simplest,
\begin{equation}
  P_{\rm TFR2}= 1,
\label{eq:ptf2}
\end{equation}
because disk galaxies define the TFR.

The fraction of bulgeless galaxies is uncertain, but they are not exceptionally rare. The quoted fraction of bulgeless among the disk galaxies varies from 10\,\% to 100\,\%, depending on the selected sample of disk galaxies, the stellar mass of the galaxy, and the redshift.  
\citet{2011ApJ...733L..47F} found that about 20\,\% of the local spirals with $M_\star\sim 10^{10}\,{\rm M_\odot}$ are bulgeless. This fraction increases to 80\,\% at $10^{9}\,{\rm M_\odot}$. 
A fraction of about 20\,\% was also found by \citet{2008ApJ...675.1194B} in a sample of some 4000 fairly luminous ($-18.5 < M_g < -22$) local galaxies extracted from the SDSS. They represent 15\,\% in the catalog of edge-on galaxies by \citet{2006A&A...445..765K}. The review by \citet{2009PASP..121.1297K} gives a range between 15\,\% and 30\,\%, depending on the type of disk.  In short, given the moderate stellar mass as a galaxy of the stellar trail ($3\times 10^{9}\,{\rm M_\odot}$; Table~\ref{tab:parameters}), the educated guess 
\begin{equation}
  P_{\galaxy bl}\gtrsim 0.2
  \label{eq:pbl}
\end{equation}
is quite reasonable.

Assuming that the galaxy disks are randomly oriented in space, the probability of finding them with an orientation $\theta$ within the range $\Delta\theta$ is $(\sin\theta/2)\,\Delta\theta$ \citep[e.g.,][]{2011ASPC..437..451S}\footnote{Derived for a random distribution of magnetic field vectors in the solar photosphere, its distribution is the same as the distribution the vectors perpendicular to  galaxy disks.}. The parameter $\theta$ is the angle between the direction perpendicular to the disk and the line of sight (Fig.~\ref{fig:edge-on-schematic}a), and therefore, edge-on views have $\theta = \pi/2$. On the other hand, a line of sight is contained within the galaxy disk provided its $\theta$ is in the interval $\pi/2\pm W_a/(2R_a)$, where $W_a$ is a representative full width of the disk, and $R_a$ is a representative radius (see Fig.~\ref{fig:edge-on-schematic}a). Thus,  $\Delta\theta\simeq W_a/R_a$, so that the probability for a disk galaxy to have an edge-on view is
\begin{equation}
  P_{\galaxy eo}\simeq  \frac{W_a}{R_a}\simeq 8\times 10^{-2},
  \label{eq:pgeo}
\end{equation}
where the numerical value corresponds to $R_a\simeq 25\,{\rm kpc}$ and $W_a=2 r_\star=2\,{\rm kpc}$.

If the mean number surface density of galaxies is $\Sigma_g$, the expected number of galaxies within a distance $D$ centered in one of them is about $\pi D^2\,\Sigma_g$. Typically, in the local Universe $\Sigma_g$ , goes from a hundred galaxies ${\rm Mpc^{-2}}$ in galaxy cluster centers to one galaxy ${\rm Mpc^{-2}}$ in the field \citep[e.g.,][]{2006A&A...445...29P,2013A&A...558A...1B,2016ApJS..224...33B}. To evaluate the number of close galaxies, we assume that they are close if  $D\lesssim 100~{\rm kpc}$, a distance representative of the size of the CGM around galaxies \citep[e.g.,][]{2013ApJ...764L..31K}. This $D$ renders between  $3\times 10^{-2}$ and 3 galaxies. Taking the number of expected nearby galaxies in the field as an upper limit of the probability of having a second galaxy close enough,
\begin{equation}
  P_{D} \gtrsim 3\times 10^{-2}.
\label{eq:pd}
\end{equation}
This inequality was derived for the local Universe, but galaxy densities are even larger at the redshift of the object.

When galaxy $a$ is viewed edge-on, the probability of the second galaxy $b$ to be aligned with the disk is about $\sqrt{W_a^2+W_b^2}/(2\pi [L-R_a])$, with $W_b$ the size of the second galaxy, and $L-R_a$ the projected distance between the two galaxies (see Fig.~\ref{fig:edge-on-schematic}b). Thus, the probability that two randomly chosen galaxies separated by $L-R_a$ look like the object  is
\begin{equation}
  P_{align}=\frac{\sqrt{W_a^2+W_b^2}}{2\pi (L-R_a)}\simeq 2.9\times 10^{-2},
\label{eq:p2nd}
\end{equation}
with the numerical value corresponding to the parameters representing the object, namely,   
$W_a=2\,{\rm kpc}$, $W_b\simeq 3W_a$, $R_a\simeq 25\,{\rm kpc}$, and $L\simeq 60\,{\rm kpc}$ (Table~\ref{tab:parameters}). 

Using Eqs.~(\ref{eq:ptf2}), (\ref{eq:pbl}),  (\ref{eq:pgeo}),  (\ref{eq:pd}),  and (\ref{eq:p2nd}), then Eq.~(\ref{eq:pgalax}) can be evaluated to render,
\begin{equation}
  P_{galax} \gtrsim 1.4\times 10^{-5}.
  \label{eq:pgalax2}
\end{equation}
Even if this probability is small, the value is more than  11 orders of magnitude larger than the probability associated with the SMBH-induced stellar wake scenario (Eq.~[\ref{eq:psmbh2}]).

\section{Discussion and conclusions}\label{sec:conclusions}

Recently,  vD23 discovered a thin long object aligned with a nearby galaxy (Fig.~\ref{fig:parameters}). It was interpreted as the stellar wake induced by the passage of an SMBH kicked out of the galaxy by the slingshot effect of a three-SMBH encounter. In principle, SMBHs can be ejected from galaxies, and the passage of a massive compact object near a molecular cloud can trigger star formation, therefore, this ingenious mechanism can potentially explain the observation. It would be the first detection of the phenomenon, and a novel pathway for the discovery of the expected runaway SMBHs \citep[e.g.,][]{2007MNRAS.377..957H,2008ApJ...677L..47D,2022arXiv221014960D,2023MNRAS.524.1987H}. However, this scenario requires several unlikely independent events to happen simultaneously. They are collected in the overall probability written down in  Eq.~(\ref{eq:psmbh}):  the probability that three SMBHs orbit each other in a galaxy, the probability that the three SMBHs are ejected from the galaxy, the probability that the galaxy has a massive radial gas filament, the probability that one of the SMBHs is shot right along the axis of the existing filament, the probability that the passage of the SMBH triggers star formation in the filament, and the probability that the formed mass is on the TFR. The last constraint is required because, as we show in Paper~I, the position-velocity curve of the stellar trail closely resembles a rotation curve, which, together with its measured stellar mass, places the object on the TFR.

Because the TFR is characteristic of galaxies, we studied the possibility that the object is a bulgeless edge-on galaxy that happens to be aligned with the second galaxy (Fig.~\ref{fig:parameters}). This is the scenario put forward in Paper~I, where the object was shown to be analogous to the well-known local edge-on galaxy IC\,5249. It was also argued that the bulgeless edge-on galaxies are not uncommon \citep{1999BSAO...47....5K}, and they are automatically on the TFR. Paralleling the calculus carried out for the SMBH runaway scenario, we estimated in Sect.~\ref{sec:alignment} the probability of this alternative possibility to explain the nature of the object. Equation~(\ref{eq:pgalax}) splits the full probability as the product of the probability for a galaxy to be bulgeless times  the probability of a disk galaxy to be observed edge-on times the probability of having a second galaxy along the major axis of the projected disk. 

The estimated probabilities are given in Eq.~(\ref{eq:pgalax2}) for the bulgeless edge-on galaxy hypothesis ($P_{galax}$), and in Eq.~(\ref{eq:psmbh2}) for  the runaway SMBH stellar trail hypothesis ($P_{\rm SMBH}$). The individual factors entering in the estimates are summarized in Table~\ref{tab:summary}. All in all, the probability of the galaxy explanation happens to exceed the probability of the runaway SMBH explanation by more than 11 orders of magnitude ($P_{galax}/P_{\rm SMBH}\sim 2\times 10^{11}$). Even acknowledging the large uncertainty of our estimates, this huge difference has to be significant.
  For example, if the runaway SMBH is produced by the recoil of a binary SMBH merger rather than by the three-body encounter, $P_{\rm SMBH}$ increases, but still $P_{galax}/P_{\rm SMBH}\sim 10^{10}$ (Appendix~\ref{app:twosmbhs}).
  Similarly, if after the three-body encounter, only one of the SMBHs is kicked out while the other two remain bound to the galaxy (a possibility disfavored by the need to conserve linear momentum; Sect.~\ref{sec:three_body}), the probability of ejection could increase by up to two orders of magnitude \citep{2007MNRAS.377..957H}, and yet $P_{galax}/P_{\rm SMBH}\sim 2\times 10^{9}$. 
Even in
the extreme case that all the six probabilities of the SMBH scenario (Eq.~[\ref{eq:psmbh}]) are underestimated by one order of magnitude each and the four entering in the galaxy scenario  (Eq.~[\ref{eq:pgalax}]\footnote{$P_{\rm TFR2}=1$ with no uncertainty.}) are overestimated by the same amount, $P_{\rm SMBH}$ would still be smaller than $P_{galax}$. A more realistic evaluation of the error in $P_{galax}/P_{\rm SMBH}$ can be carried out by invoking the central limit theorem.  The random variable $\log(P_{galax}/P_{\rm SMBH})$ is the (signed) sum of a number of independent random variables, and therefore,  according to the theorem, it approximately follows a Gaussian distribution \citep[e.g.,][]{1971stph.book.....M}. If each independent probability has an uncertainty of $\pm 0.5\,{\rm dex}$ (equivalent to an uncertainty of one order of magnitude in each probability), then the ten independent probabilities in Eqs.~(\ref{eq:psmbh}) and (\ref{eq:pgalax}) would have a combined uncertainty of $\sqrt{10}\times 0.5$, which renders
\begin{equation}
\log(P_{galax}/P_{\rm SMBH})\simeq 11.4\pm 1.6,
\end{equation}
meaning that the Gaussian random variable $\log(P_{galax}/P_{\rm SMBH})$ differs from zero by more than seven sigmas. 
Thus, even if some conditions are missing in probabilities (\ref{eq:psmbh}) and (\ref{eq:pgalax}) (e.g., the exact length of the stellar trail), those affecting $P_{\rm SMBH}$ must override those affecting  $P_{galax}$ by seven orders of magnitude for $P_{galax}$ and $P_{\rm SMBH}$ to be equal within the errors.  

The  probability of the runaway SMBH induced trail is not only small compared with the probability of the edge-on galaxy explanation. It is also negligibly small in absolute terms. In a survey with $\aleph$ galaxies, the probability of finding stellar trails caused by the SMBH mechanism is $\aleph P_{\rm SMBH}$. The large currently used  surveys have about $10^{6}$ galaxies \citep[e.g., the SDSS has $1.5\times 10^6$;][]{2011AJ....142...72E,2012ApJS..203...21A}, and so we expect $10^{-10}$ such trails. The largest surveys being carried out have $\sim~\!\!2\times 10^{8}$ galaxies \citep[e.g., DES;][]{2021ApJS..255...20A}, which predict $10^{-8}$ such trails. The next generation of surveys, represented by the Vera C. Rubin Observatory Legacy Survey of Space and Time \citep[LSST;][]{2019ApJ...873..111I}, aims to catalog $2\times 10^{10}$ galaxies, among which we expect $10^{-6}$ stellar trails created by the passage of a runaway SMBH. Finally,  the total number of galaxies  in the observed Universe up to redshift eight has been estimated to be around $2\times 10^{12}$  \citep[][]{2016ApJ...830...83C}. Even in this case, the number of predicted stellar trails is much smaller than one ($\sim 10^{-4}$).

The same exercise using the probability of the edge-on galaxy explanation (Eq.~[\ref{eq:pgalax2}]) renders a completely different picture. The number of expected coincidental alignments is about 14 for the SDSS, about 3000 for DES, about $3\times 10^5$ for the LSST, and about $3\times 10^7$ counting all the galaxies expected  in the visible Universe.
We note that the few but nonzero number of objects expected in current SDSS-like surveys is reassuring, meaning that the estimated $P_{galax}$ has the correct order of magnitude.

The numbers speak for themselves. Despite the appeal of the runaway SMBH explanation, the Occam's razor favors  that the object observed by vD23 and portrayed in Fig.~\ref{fig:parameters} is a bulgeless edge-on galaxy.
Our work does not rule out the existence of runaway SMBHs leaving stellar trails. It tells that the vD23 object is more likely to be a bulgeless edge-on galaxy than a runaway SMBH.

\begin{acknowledgements}
Thanks are due to Mireia Montes and Ignacio Trujillo for the enlightening discussions on the physics of galaxies  and on the buildup of science that eventually led to Paper~I and, later on, to this follow-up.
Thanks are also due to Sebastian F. S\'anchez  for discussions on the probability for a filament to be Jeans unstable (Eq.~[\ref{eq:punst}]),
to Laura Blecha, which rose some of the points addressed in Sect.~\ref{app:appa}, and to Ra\'ul de la Fuente Marcos, for clarifications on his paper.
  I am also grateful to an anonymous referee for several constructive ideas to complete the argumentation and, in particular, 
for suggesting considering the scenario with a binary SMBH merger (Appendix~\ref{app:twosmbhs}).   
This work has been partly supported by   the Spanish Ministry of Science and Innovation through projects PID2019-107408GB-C43 and PID2022-136598NB-C31 (ESTALLIDOS).
\end{acknowledgements}

%

\appendix
  \section{Runaway SMBHs produced by binary SMBH mergers}\label{app:twosmbhs}
  Galaxy mergers produce binary SMBHs that also tend to merge sometime after the galaxy merger.  The recoil induced by the anisotropic emission of gravitational waves during the  SMBH merger occasionally  leads to the ejection of the newly formed SMBH \citep[e.g.,][]{2007PhRvL..98w1102C,2012PhRvD..85h4015L}. This other channel for creating runaway SMBHs is also mentioned as a surrogate alternative by vD23. In this case, the total probability is similar to the probability we determined in Sect.~\ref{sec:smbh_scenario}, except that Eq.~(\ref{eq:psmbh}) has to be replaced with 
\begin{equation}
  P_{\rm SMBH}=P_{\rm 2\bullet}\,P_{eject}\,P_{fila}\,P_{shot}\,P_{unst}\,P_{\rm TFR1},
  \label{eq:twosmbhs}
\end{equation}
where $P_{\rm 2\bullet}$ is the probability that a randomly chosen galaxy has a binary SMBH that is ready to merge, and 
$P_{eject}$ is the probability that the fused SMBH receives a final kick that is strong enough to become a runaway SMBH. The other probabilities appearing in Eq.~(\ref{eq:twosmbhs}) are identical to those in Eq.~(\ref{eq:psmbh}) and  Table~\ref{tab:summary}. The estimate of $P_{\rm 2\bullet}$ is quite uncertain, but we can set an upper limit replicating the estimate of $P_{\rm 3\bullet}$ made in the main text. If the typical lifetime for the close SMBH binary is about 0.01\,--\,0.5\,Gyr \citep[e.g.,][]{2016ApJ...828...73K,2018MNRAS.473.3410R,2022ApJ...929..167M}, and the galaxies go through major mergers only a few times during the Hubble time (e.g., three times), this leads to a probability
\begin{equation}
  P_{\rm 2\bullet} < \frac{3\times 0.25\,{\rm Gyr}}{t_H}\simeq 5.6\times 10^{-2},
  \label{eq:pof2}
\end{equation}
which is only a factor of two larger than the upper limit used for  $P_{\rm 3\bullet}$ in the main text (Eq.~[\ref{eq:pof3}]).
The value for $P_{eject}$ is also extremely uncertain, but as happened with the three-SMBH scenario, it has to be small. Most of the mergers produce SMBH with recoil velocities well below the escape velocity. The actual recoil depends on the alignment of the spins with respect to the orbital plane and can reach thousands of km\,s$^{-1}$ when both spins lie in this plane \citep[e.g.,][]{2007PhRvL..98w1102C}. This configuration is unlikely, however. Whether the configuration is favored depends very much on the interaction of the two SMBHs with the circumbinary accretion disk \citep[e.g.,][]{2015MNRAS.451.3941G}.  Numerical simulations of the process by  \citet[][]{2012PhRvD..85h4015L} predicted recoils of about 2000~km\,s$^{-1}$ (as needed to explain vD23 scenario) with a probability from $4\times 10^{-4}$ to $10^{-2}$ , depending on the unknown details of the accretion disk. The initial conditions of the two SMBHs drastically change the calculated recoil; for equal-mass binaries and randomly oriented spins, \citet{2018PhRvD..97j4049G} reported kicks in excess of  2000~km\,s$^{-1}$ in 2.5\,\% of the cases.
Similarly, a uniform distribution of spin directions renders probabilities between $5\times 10^{-6}$ and $1.5\times 10^{-2}$ \citep{2013PhRvD..87h4027L}. Thus, all uncertainties notwithstanding, we use as a conservative upper limit for ejection a value of 
\begin{equation}
    P_{eject} \lesssim 10^{-2},
    \label{eq:peject2}
\end{equation}
which is one order of magnitude larger than the probability associated with the three-body encounters  (Eq.~[\ref{eq:peject}]).

  With the probabilities in Eqs.~(\ref{eq:pof2}) and (\ref{eq:peject2}) and the remaining probabilities from Table~\ref{tab:summary}, the likelihood of the SMBH event (Eq.~[\ref{eq:twosmbhs}]) becomes 
\begin{equation}
P_{\rm SMBH} < 1.3\times 10^{-15}.
  \label{eq:psmbh2_2}
\end{equation}
This is 20 times larger than the estimate based on three SMBH encounters (Eq.~[\ref{eq:psmbh2}]) but still exceedingly small, and it is fully consistent with the conclusions reached in Sect.~\ref{sec:conclusions}. Namely, it is much smaller than the edge-on galaxy probability (Eq.~[\ref{eq:pgalax2}]),  and the chances of finding one such runaway SMBH remain negligible.

\end{document}